\documentstyle[11pt,paspconf,epsfig]{article}

\begin{document}

\title{Galaxy-galaxy lensing in clusters: new results}

\author{Priyamvada Natarajan$^{1,2}$, Jean-Paul Kneib$^{3}$ \& Ian Smail$^{4}$}
\affil{1 Institute of Astronomy, Madingley Road, Cambridge CB3 0HE, UK}
\affil{2 Department of Astronomy, Yale University, New Haven, CT, USA}
\affil{3 Observatoire Midi-Pyrenees, 14 Av. E.Belin, 31400 Toulouse,
France}
\affil{4 University of Durham, Department of Physics, South Road, 
Durham DH1 3LE, UK}

\begin{abstract}
A synopsis of the recent results from our study of galaxy-galaxy
lensing in clusters is presented. We have applied our analysis
techniques to the sample of {\it HST} cluster-lenses that span a
redshift range from $z = 0.18$--0.58. We find that there is evidence
that the total mass of a typical early-type cluster L$^*$ galaxy
increases with redshift. For the lowest redshift bin, a sensible
comparison can be made with field galaxies and it is found that
cluster galaxies are less massive and less extended than equivalent
luminosity field galaxies. This agrees broadly with the theoretical
picture of tidal stripping of dark matter halos in high density
cluster environments. We also find the following trends -- at a fixed
luminosity both the mass-to-light ratio and the fiducial truncation
radius increase with redshift from $z = 0.18$--0.58.
\end{abstract}

\keywords{gravitational lensing, galaxies: fundamental parameters, 
halos, methods: numerical}

\section{Introduction}

Recent work on galaxy-galaxy lensing in the cores of rich clusters
suggests that the average mass-to-light ratio and spatial extents of
the dark matter halos associated with morphologically-classified
early-type cluster members is significantly different from those of
comparable luminosity field galaxies (Natarajan et al.\ 1998
[N98]). For field galaxies galaxy-galaxy lensing has been used to
place more uncertain constraints on halo masses and sizes, with the
claim that halos of field galaxies extend beyond 100 kpc (Brainerd,
Blandford \& Smail 1996; Ebbels et al.\ 2000; Hudson et al.\ 1998).

The detailed mass distribution within clusters -- the fraction of the
total cluster mass that is associated with individual galaxies -- has
important consequences for the frequency of galaxy interactions.  The
global tidal field of the cluster potential well is strong enough to
truncate the dark matter halo of a galaxy whose orbit penetrates the
cluster core. Compact dark halos indicate a high probability for
galaxy--galaxy collisions over a Hubble time within a rich
cluster. However, since the internal velocity dispersions of cluster
galaxies ($\sim 120$--200 km s$^{-1}$) are significantly lower than
their orbital velocities, these interactions are in general unlikely
to lead to mergers, suggesting that the encounters of the kind
simulated in the galaxy harassment picture by Moore et al.\ (1996) are
frequent and lead to morphological transformation. In high resolution
N-body simulations of galaxy halos within a rich cluster Ghigna et
al.\ (1998) report that halos that traverse within the inner $200$ kpc
of the cluster centre suffer significant tidal truncation.

~From the analysis of local weak distortions in the cluster AC114 at a
redshift of 0.31, N98 found that the total mass of a fiducial $L^*$
cluster spheroidal galaxy was contained within $\sim$ 15 kpc radius
halo ($\sim$~8--10 $R_e$) with a mass-to-light ratio
${M/L_V}\,\sim\,{23^{+15}_{-6}} $ (90 \% c.l., $h=0.5$) in solar units
within this radius. This limit on the truncation radius points to
cluster galaxies having of a significantly more compact and less
massive halo than an equivalently luminous field galaxy.

However, we point out that several complex biases need to be taken
into account to extend the analysis to such a
non-uniform sample -- these HST cluster-lenses span a large
range in mass, richness, and X-ray luminosity, fortunately, they form a
subset of the well-studied MORPHS clusters (Couch et al.\ 1998; Smail
et al.\ 1997).  No significant evolution in luminosity was found in
the spheroidal populations of these clusters (Barger et al.\ 1998),
therefore, any likely biases arising purely from differences in star
formation activity due to different morphological mixes at various
redshifts are expected to be small.

\section{Analysis techniques}

We model the cluster potential as a composite of a large-scale smooth
component and the sum of smaller-scale perturbers which are associated
with bright, early-type galaxies in the cluster. Details of the
procedure can be found elsewhere (Natarajan \& Kneib 1997; N98). Using
both the observed strong lensing features and the shear field as
constraints, the relative fraction of mass that can be attributed to
the smooth component and the perturbers are computed using a
maximum-likelihood method. The likelihood prescription provides bounds
on both the fiducial central velocity dispersion (in km/s) and the
outer extent (in kpc) for an ensemble of galaxies using a
parameterised description of the scaling of the velocity dispersion
and truncation radius with luminosity (the constraints are not
sensitive to the details of this parametric form).

\section{Results and Conclusions}

Here we present the results of the application of our technique to the
WFPC2 images of five {\it HST} cluster-lenses (Couch et al.\ 1998;
Smail et al.\ 1997). The clusters span a wide redshift range: A\,2218
at $z = 0.18$; AC\,114 at $z = 0.31$; Cl\,0412$-$65 at $z = 0.51$;
Cl\,0016+16 at $z = 0.55$ and Cl\,0054$-$27 at $z = 0.58$.  We find
that the mass-to-light ratio in the $V$-band of a typical L$^*$
increases as a function of redshift, with a mean value that ranges
from roughly 10--24.  The average value of the fiducial truncation
radius varies from about 15 kpc at $z = 0.18$ to 60 kpc at $z =
0.58$. For the galaxy models that we have used in our analysis, the
typical total mass of an L$^*$ varies with redshift from $\sim 2.8
\times 10^{11}\,M_{\odot}$ to $\sim 7.7 \times 10^{11}\,M_{\odot}$.
The mass-to-light ratios quoted here take passive evolution of
elliptical galaxies into account as given by the stellar population
synthesis models of Bruzual \& Charlot. The mass obtained for a
typical bright cluster galaxy by Tyson et al.\ (1998) from the strong
lensing analysis of the cluster Cl\,0024+16, at $z = 0.41$, is
consistent with our results.  Note that the preliminary numbers quoted
here are average values, the estimates and discussion of error bars is
presented in Natarajan, Kneib \& Smail (2000).
 
These observed trends are in good agreement with theoretical
expectations from cluster formation and evolution models (Ghigna et
al.\ 1998). Qualitatively, in the context of this paradigm, fewer
cluster members in a dynamically younger cluster of galaxies are
likely to have suffered passages through the inner regions and are
hence expected to be less tidally stripped. Detailed analysis and
interpretation of these results are presented in a forthcoming paper.


\begin{references}
\reference Barger, A. J., et al., 1998, \apj, 501, 522

\reference Brainerd, T., Blandford, R., \& Smail, I., 1996, \apj, 466, 623
 
\reference Couch, W.J., Barger, A.J., Smail, I., Ellis, R.S., Sharples,
R.M., 1998, ApJ, 497, 188

\reference Ebbels, T. M. D. et al., 2000, in preparation.

\reference Ghigna, S., Moore, B., Governato, F., Lake, G., 
Quinn, T., \& Stadel, J., 1998, \mnras, 300, 146

\reference Hudson, M. J., Gwyn, S. D. J., Dahle, H., \& Kaiser, N.,
1998, \apj, 503, 531

\reference Moore, B., Katz, N.,  Lake, G., Dressler, A., \& 
Oemler, A., 1996, Nature, 379, 613

\reference Natarajan, P., \& Kneib, J-P., 1997, \mnras, 287, 833 

\reference Natarajan, P., Kneib, J-P., Smail, I., 
\& Ellis, R. S., 1998, \apj, 499, 600  [N98] 

\reference Natarajan, P., Kneib, J-P., \& Smail, I., 2000, in 
preparation

\reference Smail, I., Dressler, A., Couch, W.J., Ellis, R.S., Oemler,
A., Butcher, H., Sharples, R.M., 1997, ApJS, 110, 213  

\reference Tyson, J. A., Kochanski, G., \& De'll Antonio, I. P., 
1998, \apj, 498, L107
 \end{references}
\end{document}